\documentclass[prc,twocolumn,superscriptaddress,twoside,floatfix]{revtex4-2}

\usepackage{amssymb,epsfig}
\usepackage{xcolor}

\usepackage{graphicx}
\usepackage[outdir=./]{epstopdf}

\begin{document}

\title{Resonant states in $^{7}$H. I. Experimental studies of the
$^2$H$(^8$He$,^3$He) reaction}

\author{I.A.~Muzalevskii}
\email{muzalevsky@jinr.ru}
\affiliation{Flerov Laboratory of Nuclear Reactions, JINR,  141980 Dubna,
Russia}
\affiliation{Institute of Physics, Silesian University in Opava, 74601  Opava,
Czech Republic}

\author{A.A.~Bezbakh}
\affiliation{Flerov Laboratory of Nuclear Reactions, JINR,  141980 Dubna,
Russia}
\affiliation{Institute of Physics, Silesian University in Opava, 74601 Opava,
Czech Republic}

\author{E.Yu.~Nikolskii}
\affiliation{National Research Centre ``Kurchatov Institute'', Kurchatov sq.\ 1,
123182 Moscow, Russia}
\affiliation{Flerov Laboratory of Nuclear Reactions, JINR,  141980 Dubna,
Russia}

\author{V.~Chudoba}
\affiliation{Flerov Laboratory of Nuclear Reactions, JINR,  141980 Dubna,
Russia}
\affiliation{Institute of Physics, Silesian University in Opava, 74601  Opava,
Czech Republic}

\author{S.A.~Krupko}
\affiliation{Flerov Laboratory of Nuclear Reactions, JINR,  141980 Dubna,
Russia}

\author{S.G.~Belogurov}
\affiliation{Flerov Laboratory of Nuclear Reactions, JINR,  141980 Dubna,
Russia}
\affiliation{National Research Nuclear University ``MEPhI'', 115409 Moscow,
Russia}

\author{D.~Biare}
\affiliation{Flerov Laboratory of Nuclear Reactions, JINR,  141980 Dubna,
Russia}

\author{A.S.~Fomichev}
\affiliation{Flerov Laboratory of Nuclear Reactions, JINR,  141980 Dubna,
Russia}
\affiliation{Dubna State University, 141982 Dubna, Russia}

\author{E.M.~Gazeeva}
\affiliation{Flerov Laboratory of Nuclear Reactions, JINR,  141980 Dubna,
Russia}

\author{A.V.~Gorshkov}
\affiliation{Flerov Laboratory of Nuclear Reactions, JINR,  141980 Dubna,
Russia}

\author{L.V.~Grigorenko}
\affiliation{Flerov Laboratory of Nuclear Reactions, JINR,  141980 Dubna,
Russia}
\affiliation{National Research Nuclear University ``MEPhI'', 115409 Moscow,
Russia}
\affiliation{National Research Centre ``Kurchatov Institute'', Kurchatov sq.\ 1,
123182 Moscow, Russia}

\author{G.~Kaminski}
\affiliation{Flerov Laboratory of Nuclear Reactions, JINR,  141980 Dubna,
Russia}
\affiliation{Heavy Ion Laboratory, University of Warsaw, 02-093 Warsaw, Poland}

\author{O.~Kiselev}
\affiliation{GSI Helmholtzzentrum f\"ur Schwerionenforschung GmbH, 64291
Darmstadt, Germany}

\author{D.A.~Kostyleva}
\affiliation{GSI Helmholtzzentrum f\"ur Schwerionenforschung GmbH, 64291
Darmstadt, Germany}
\affiliation{II. Physikalisches Institut, Justus-Liebig-Universit\"at, 35392
Giessen, Germany}

\author{M.Yu.~Kozlov}
\affiliation{Laboratory of Information Technologies, JINR,  141980 Dubna,
Russia}

\author{B. Mauyey}
\affiliation{Flerov Laboratory of Nuclear Reactions, JINR,  141980 Dubna,
Russia}
\affiliation{Institute of Nuclear Physics, 050032 Almaty, Kazakhstan}

\author{I.~Mukha}
\affiliation{GSI Helmholtzzentrum f\"ur Schwerionenforschung GmbH, 64291
Darmstadt, Germany}

\author{Yu.L.~Parfenova}
\affiliation{Flerov Laboratory of Nuclear Reactions, JINR,  141980 Dubna,
Russia}

\author{W.~Piatek}
\affiliation{Flerov Laboratory of Nuclear Reactions, JINR,  141980 Dubna,
Russia}
\affiliation{Heavy Ion Laboratory, University of Warsaw, 02-093 Warsaw, Poland}

\author{A.M.~Quynh}
\affiliation{Flerov Laboratory of Nuclear Reactions, JINR,  141980 Dubna,
Russia}
\affiliation{Nuclear Research Institute, 670000 Dalat, Vietnam}

\author{V.N.~Schetinin}
\affiliation{Laboratory of Information Technologies, JINR,  141980 Dubna,
Russia}

\author{A.~Serikov}
\affiliation{Flerov Laboratory of Nuclear Reactions, JINR,  141980 Dubna,
Russia}

\author{S.I.~Sidorchuk}
\affiliation{Flerov Laboratory of Nuclear Reactions, JINR,  141980 Dubna,
Russia}

\author{P.G.~Sharov}
\affiliation{Flerov Laboratory of Nuclear Reactions, JINR,  141980 Dubna,
Russia}
\affiliation{Institute of Physics, Silesian University in Opava, 74601 Opava,
Czech Republic}

\author{N.B.~Shulgina}
\affiliation{National Research Centre ``Kurchatov Institute'', Kurchatov sq.\ 1,
123182 Moscow, Russia}
\affiliation{Bogoliubov Laboratory of Theoretical Physics, JINR, 141980 Dubna,
Russia}

\author{R.S.~Slepnev}
\affiliation{Flerov Laboratory of Nuclear Reactions, JINR,  141980 Dubna,
Russia}

\author{S.V.~Stepantsov}
\affiliation{Flerov Laboratory of Nuclear Reactions, JINR,  141980 Dubna,
Russia}

\author{A.~Swiercz}
\affiliation{Flerov Laboratory of Nuclear Reactions, JINR,  141980 Dubna,
Russia}
\affiliation{AGH University of Science and Technology, Faculty of Physics and
Applied Computer Science, 30-059 Krakow, Poland}

\author{P.~Szymkiewicz}
\affiliation{Flerov Laboratory of Nuclear Reactions, JINR,  141980 Dubna,
Russia}
\affiliation{AGH University of Science and Technology, Faculty of Physics and
Applied Computer Science, 30-059 Krakow, Poland}

\author{G.M.~Ter-Akopian}
\affiliation{Flerov Laboratory of Nuclear Reactions, JINR,  141980 Dubna,
Russia}
\affiliation{Dubna State University, 141982 Dubna, Russia}

\author{R.~Wolski}
\affiliation{Flerov Laboratory of Nuclear Reactions, JINR,  141980 Dubna,
Russia}
\affiliation{Institute of Nuclear Physics PAN, Radzikowskiego 152, 31342
Krak\'{o}w, Poland}

\author{B.~Zalewski}
\affiliation{Flerov Laboratory of Nuclear Reactions, JINR,  141980 Dubna,
Russia}
\affiliation{Heavy Ion Laboratory, University of Warsaw, 02-093 Warsaw, Poland}

\author{M.V.~Zhukov}
\affiliation{Department of Physics, Chalmers University of Technology, S-41296
G\"oteborg, Sweden}

\date{\today.}

\begin{abstract}
The extremely neutron-rich system $^{7}$H was studied in the direct
$^2$H($^8$He,$^3$He)$^7$H transfer reaction with a 26\,AMeV secondary $^{8}$He
beam \cite{Bezbakh:2020}.
The missing mass spectrum and center-of-mass (c.m.) angular distributions of
$^{7}$H, as well as the momentum distribution of the $^{3}$H fragment in the
$^{7}$H frame, were constructed.
In addition to the investigation reported in Ref.\ \cite{Bezbakh:2020}, we
carried out another experiment with the same beam but a modified setup, which
was cross-checked by the study of the $^2$H($^{10}$Be,$^3$He$)^{9}$Li reaction.
A solid experimental evidence is provided that two resonant states of $^{7}$H
are located in its spectrum at $2.2(5)$ and $5.5(3)$\,MeV relative to the
$^3$H+4$n$ decay threshold.
Also, there are indications that the resonant states at $7.5(3)$ and
$11.0(3)$\,MeV are present in the measured $^{7}$H spectrum.
Based on the energy and angular distributions, obtained for the studied
$^2$H($^8$He,$^3$He)$^7$H reaction, the  weakly populated $2.2(5)$\,MeV peak is
ascribed to the $^7$H ground state.
It is highly plausible that the firmly ascertained 5.5(3)\,MeV state is the
$5/2^+$ member of the $^7$H excitation  $5/2^+$--$3/2^+$ doublet, built on the
$2^+$ configuration of valence neutrons.
The supposed $7.5$\,MeV state can be another member of this doublet, which could
not be resolved in Ref.\ \cite{Bezbakh:2020}.
Consequently, the two doublet members appeared in the spectrum of $^{7}$H in
\cite{Bezbakh:2020} as a single broad $6.5$\,MeV peak.
\end{abstract}

\maketitle


\section{Introduction}


Exploration of exotic nuclei located in the vicinity of the neutron drip line
has led to several remarkable discoveries, such as the neutron haloes and skins,
shell quenching, and the appearance of new magic numbers.
It also belongs to the modern trends to investigate the most neutron-rich
systems as far beyond the dripline as possible.
Among the recent results in this field one should mention the works on $^{10}$He
\cite{Sidorchuk:2012,Kohley:2012,Jones:2015,Matta:2015}, $^{13}$Li
\cite{Johansson:2010a,Kohley:2013b}, $^{16}$Be \cite{Spyrou:2012}, $^{21}$B
\cite{Leblond:2018}, $^{26}$O \cite{Kohley:2013,Caesar:2013,Kondo:2016} and the
on-going quest for $^{18}$Be, $^{28}$O, $^{33}$F \cite{Ahn:2019} (and analogous
very exotic species).
One can find that these expements require extreme efforts, often leadinging to
poor data quality (statistics, resolution) and, in turn, to numerous unresolved
questions and controversies, see, e.g.\ \cite{Grigorenko:2016,Fortune:2018}.
The typical feature of the mentioned nuclides is the multineutron (at least
two-neutron) emission and $^{7}$H with its four-neutron decay channel represents
a very important guideline case for the prospective studies in this field.

\begin{figure}
\begin{center}
\includegraphics[width=0.45\textwidth]{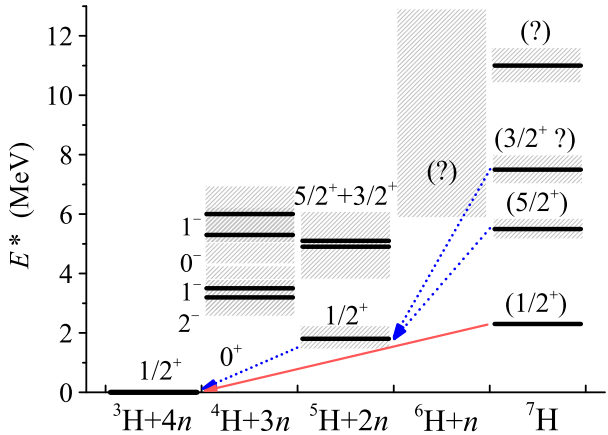}
\end{center}
\caption{
The $^{7}$H level scheme, and the known neighboring $^{5}$H
\cite{Korsheninnikov:2001,Golovkov:2004a,Golovkov:2005} and $^{4}$H systems
important for the discussions of this work.
The solid red arrow illustrates the decay mechanism of the $^{7}$H ground state
(g.s.) which is expected to be ``true'' $4n$ emission.
The dotted blue arrows illustrate the decay mechanism of the higher excitations
in $^{7}$H, which is expected to be the sequential $(2n)+(2n)$ emission via the
$^{5}$H g.s.
}
\label{fig:levels}
\end{figure}

Going beyond the neutron drip line we enter the region where the conditions of
``true'' $4n$ emission are probably valid for some nuclei.
The attribute ``true'', applied in particular to the $4n$ decay, emphasizes the
absence of sequential neutron emission, and that the decay is possible only via
the five-body core+$4n$ simultaneous emission, see the illustration of the
$^{7}$H case in Fig.\ \ref{fig:levels}.
The true $4n$ decay of $^{7}$H, $^{28}$O, and some other nuclei with enormous
neutron excess was the subject of detailed consideration in Ref.\
\cite{Grigorenko:2011}.
By applying the formalism based on the simplified three-body and five-body
Hamiltonians, the authors showed that the few-body dynamics of the $2n$ and $4n$
emission leads to collective barriers which increase rapidly with the increase
of number of emitted particles.
Therefore, the prospects to detect extremely long-lived resonances open for true
$4n$ decay seem to be more promising than for $2n$ decay.
The discovery of the so far unexplored phenomenon of true $4n$ emission is a
task of fundamental importance.
The $^{7}$H nucleus is evidently a suitable candidate for the outlined
investigations.

In this work, we use the $E_T$ notation for the decay energy above the
corresponding threshold, e.g., $^{3}$H+$4n$ for $^{7}$H or $^{3}$H+$2n$ for
$^{5}$H. The experimentally determined missing mass (MM) energy is calibrated in
the same way.


\subsection{History of the research subject}


The first theoretical estimations of Baz' and coworkers \cite{Baz:1972}
predicted that the $^{7}$H nucleus could be bound.
However, the experiments \cite{Seth:1981,Evseev:1981} searching for $^{7}$H
formed in the $^7$Li($\pi^-,\pi^+$) reaction gave negative results.
Also, the experiment \cite{Aleksandrov:1982} aimed to detect this nucleus among
the ternary fission products of $^{252}$Cf provided no evidence.
The observation of the ground state resonance in $^{5}$H
\cite{Korsheninnikov:2001} revived theoretical interest to the possible
existence of a low-lying $^{7}$H  state near the $^{3}$H+$4n$ decay threshold.
Calculations using the seven-body hyperspherical functions formalism
\cite{Timofeyuk:2002} evaluated the $^{7}$H g.s.\ energy as $E_{T} \approx
3$\,MeV.
In Ref.\ \cite{Korsheninnikov:2003} the binding energy of the $^{7}$H ground
state was estimated to be $\sim 5.4$\,MeV, which means that this resonance state
is expected at about 3\,MeV above the $^3$H+$4n$ decay threshold.
The authors emphasized that the $^{7}$H ground state should undergo the unique
five-body decay into $^3$H+$4n$ with very small width. The phenomenological
estimates in Ref.\ \cite{Golovkov:2004} pointed to $E_{T} \sim 1.3-1.8$\,MeV.
The calculations within antisymmetrized molecular dynamics \cite{Aoyama:2004}
and \cite{Aoyama:2009} provided $E_{T} \sim 7$ and $E_{T} \sim 4$\,MeV,
respectively.

The first experimental evidence of the $^{7}$H g.s.\ resonance was observed in
the study of the $^1$H($^8$He,$2p)^7$H reaction in Ref.\
\cite{Korsheninnikov:2003}.
The MM spectrum of $^{7}$H obtained in that work showed a sharp increase
starting from the $^3$H+$4n$ threshold.
Nevertheless, this interesting observation did not allow the authors to give
quantitative information about the resonance parameters because of low energy
resolution (of $\sim 2$\,MeV) and complicated background conditions.

A sophisticated approach was used in the work \cite{Golovkov:2004} carried out
by the ACCULINNA fragment-separator group.
By bombarding a very thick (5.6 cm) liquid deuterium target with a beam of
20.6\,AMeV $^{8}$He  projectiles, the authors searched for the quasistable
$^{7}$H nuclei produced in the $^2$H($^8$He,$^7$H)$^3$He reaction within
$0^{\circ}-50^{\circ}$ c.m.\ angular range and with such a lifetime longer than
1\,ns.
No $^7$H events with such lifetime was found.
This gives a very low limit for the cross section of the
$^2$H($^8$He,$^7$H)$^3$He reaction, $\sigma < 3$ nb/sr, which is by several
orders of the magnitude less than the expected value.
The lifetime estimates made in Ref.\ \cite{Golovkov:2004} led to the conclusion
that the obtained limit of the $^{7}$H production cross section implies a lower
limit of $E_T \gtrsim 50-100$ keV for its decay energy.
This indicates that the only realistic approach to the $^7$H problem is the
search for the shorter-lived resonance states of this nucleus in the five-body
$^{3}$H+$4n$ continuum.

Results obtained in the study of stopped $\pi^-$ absorption by the $^9$Be and
$^{11}$B targets were reported in Ref.\ \cite{Gurov:2007}.
The count rate of the $p$+$^3$He products emitted in the
$^{11}$B($\pi^-$,$p^3$He)$^7$H reaction was very low.
The authors concluded that the question of the possible existence of the $^{7}$H
states, both near the $^3$H+$4n$ threshold and in the region of higher
excitation energy remains open \cite{Gurov:2009}.

The $^{7}$H existence was investigated by the authors of Refs.\
\cite{Caamano:2007,Caamano:2008} in the transfer reaction
$^{12}$C($^{8}$He,$^{13}$N)$^{7}$H.
Although in this work only seven events could be attributed to the desired
reaction channel, a very narrow $^7$H resonance was announced, with $E_T=
0.57^{+0.42}_{-0.21}$\,MeV.
It should be pointed out that no actual reaction channel identification was
possible in this experiment.
The interpretation is essentially based on the assumption that only the $^{7}$H
g.s.\ is populated in this reaction.
In reality, the population of $^{7}$H$^*$ is also possible in this experiment.
In addition, the reactions $^{12}$C($^{8}$He,$^{14}$N)$^{6}$H and
$^{12}$C($^{8}$He,$^{15}$N)$^{5}$H may mock up the detection of $^{7}$H.

The authors of Ref.\ \cite{Fortier:2007} investigated the
$^2$H($^8$He,$^3$He)$^7$H reaction.
They concluded that there was some indication of a $^7$H resonance state in the
measured MM spectrum at $E_T \sim 2$\,MeV.
It is notable, however, that the experimental acceptance covered only the
energies up to 5\,MeV in the $^{7}$H excitation spectrum.
Within this narrow energy window, the $^{7}$H spectrum from the
$^2$H($^8$He,$^3$He)$^7$H reaction looks very similar to the spectrum of the
carbon-induced background from the CD$_2$ target, which made the authors
cautious about their observations.

The next attempt to discover $^{7}$H  using the $^2$H($^8$He,$^3$He)$^7$H
reaction was made in Ref.\ \cite{Nikolskii:2010} at RIKEN.
No indication on the resonance peak was revealed in the measured $^7$H MM
spectrum.
However, some peculiarity was found in this spectrum at $\sim 2$\,MeV above the
$^3$H+$4n$ decay threshold.
The authors reported a value of about 30 $\mu$b/sr in c.m.\ for the
cross-section of the reaction populating the low-energy part in the $^7$H
spectrum.
In addition, they noted that the $^7$H spectrum demonstrates a peculiarity at
about 10.5\,MeV that could be a manifestation of a $^{7}$H continuum excitation.


\subsection{Presented experimental results}


One may conclude, that the best approach to studies of the unstable $^{7}$H
nucleus suggests the use of the $^2$H($^8$He,$^3$He)$^7$H reaction.
But the low intensity of the $^{8}$He beam produced at the ACCULINNA separator
allowed us only to put a limit of the cross section of the
$^2$H($^8$He,$^3$He)$^7$H reaction near the $^3$H+$4n$ decay threshold
\cite{TerAkopian:2007}.
The new radioactive-beam separator ACCULINNA-2 commissioned at the Flerov
Laboratory of Nuclear Reactions (FLNR) in 2017 \cite{Fomichev:2018} provides the
$^{8}$He beam with intensity up to $\sim 10^5$ pps, which is sufficient for the
challenging experiments aimed at $^{7}$H.

The first results of our studies of the $^{7}$H spectrum were published as a
Letter \cite{Bezbakh:2020}. We confidently observed the resonant structure at
$E_T = 6.5(5)$\,MeV interpreted as an overlapping doublet of $3/2^+$ and $5/2^+$
states (this is the $2^+$ excitation of the valence nucleons coupled with the
$1/2^+$ spin-parity of the $^{3}$H core).
There was a group of events at $\sim 2 $\,MeV which was considered as a
candidate for the $^{7}$H $1/2^+$ ground state with $E_T = 1.8(5)$\,MeV.
However, due to the low statistics (5 events), there was no complete confidence
in such an interpretation.
The estimated cross section of the reaction channel populating this possible
state appeared to be quite low.
The value $d \sigma /d \Omega \sim 25$ $\mu$b/sr was derived from the five
$^{7}$H g.s.\ counts detected in the c.m.\ angular range $19^{\circ} -
27^{\circ}$.
The bump, present at $E_T > 10$\,MeV in the spectrum, was fitted by assuming a
resonance contribution at $E_T = 12$\,MeV with $\Gamma = 4$\,MeV.
Such an interpretation was quite cautious since this bump is close to the
experimental cutoff of the measured $^{7}$H MM spectrum.

In the present work we further elaborate the data analysis of
\cite{Bezbakh:2020} (``experiment 1'').
We also present the data of a new experiment (``experiment 2'') performed with
the same beam, but with an improved setup.
We present new data inferring more information about the excited resonant state
in $^{7}$H \cite{Bezbakh:2020} and to get clear results that would reliably
characterize the $^{7}$H ground state.
In particular, we extend the measured spectrum to the smaller c.m.\ angular
range.
The accumulated number of $^{7}$H events in the new dataset is more than three
times larger, than in the first run.
The calibration of the $^{7}$H MM spectrum is independently verified by the
$^2$H($^{10}$Be,$^3$He)$^{9}$Li reaction carried out in a dedicated experiment
with $^{10}$Be beam.
The new data confirms the spectrum reported in Ref.\ \cite{Bezbakh:2020}.
Based on the theoretical estimations and Monte-Carlo (MC) simulations, provided
in this and in the forthcoming work \cite{Grigorenko:2021}, we provide a solid
experimental evidence of the population of the resonant states in $^{7}$H at
$2.2(5)$ and $5.5(3)$\,MeV.
There is also some evidence of the resonance states at $7.5(3)$ and
$11.0(3)$\,MeV.


\section{Experimental setup}
\label{sec:exp}


The experiments were carried out at the FLNR, JINR, with the use of Radioactive
Ion Beam (RIB) produced by the ACCULINNA-2 fragment separator.
The primary beam of $^{11}$B ($\sim 1$ p$\mu$A, 33.4\,AMeV) or $^{15}$N ($\sim
0.5$ p$\mu$A, 49.7\,AMeV) ions accelerated by the U-400M cyclotron bombarded the
1\,mm thick beryllium production target installed at the initial focal plane of
the separator.
As a result of fragmentation and subsequent separation, the $^8$He and $^{10}$Be
beams were obtained with intensity $\sim 10^5$ pps, and their energies in the
middle of D$_2$ target spread within $\pm 7 \%$ and $\pm 2.5 \%$ around the mean
values of 26 and 42\,AMeV, respectively.

\begin{figure}
\begin{center}
\includegraphics[width=0.47\textwidth]{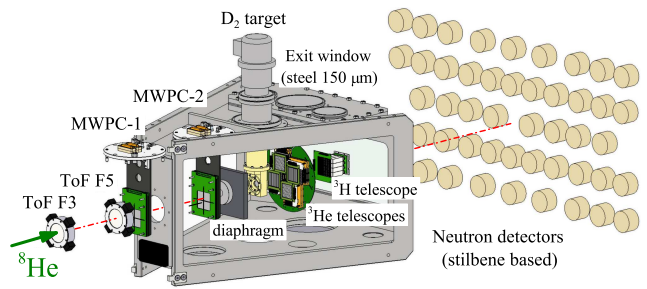}
\end{center}
\caption{
Experimental setup common for the experiments 1 and 2 at the final focal plane
F5 of the ACCULINNA-2 fragment separator.
Detectors of $^8$He projectile positions (MWPC-1,2) and time-of-flight (ToF) are
described in the text.
There is no common scale along the beam axis in this plot.
The telescope detectors of the reaction products $^3$H and $^3$He are also shown
in Figs.\ \ref{fig:setup-1} and \ref{fig:setup-2}.
}
\label{fig:exp-common}
\end{figure}

The sketch of the setup is shown in Fig.\ \ref{fig:exp-common}.
The energy values of the individual RIB projectiles were determined with
precision of $ \sim 0.2 \%$ by means of the Time-of-Flight (ToF) detector.
The two BC404 plastic scintillators placed at a ToF base of 12.3 meters allowed
for identification of the RIB projectiles by the $\Delta E$-ToF method
\cite{Kaminski:2020}.
Both RIB's, $^8$He and $^{10}$Be, were well separated with purities better than
$90\%$ and $80\%$, respectively.

The beam tracking was arranged by a pair of the Multi-Wire Proportional Chambers
(MWPC) placed at the distances of 28 and 81\,cm upstream of the gaseous target.
This allowed for determination of the RIB interaction points in the target plane
with a 1.8\,mm  precision.
Also, using this beam-tracking installation we determined the inclination angles
of individual RIB projectiles to the ion optical axis with an accuracy of $\sim
0.15 $\,degrees.

The 4\,mm thick target cell, equipped with the 6\,$\mu$m thick and 25\,mm
diameter stainless-steel entrance and exit windows, was cooled down to 27\,K and
filled with the deuterium gas up to a pressure at which the target thickness was
$\sim 3.7 \times 10 ^{20}$ cm$^{-2}$.
The cell was concealed in a screened volume having a pair of 3.5\,$\mu$m thick
aluminum-backed Mylar windows and kept cooled to the same temperature to ensure
thermal protection.
The entrance/exit target windows, deformed by the gas pressure, took the
near-lenticular form, so that the maximum target thickness turned out to be
6\,mm.

The part of the experimental setup described above was common for both
experiments investigating the $^7$H MM spectrum populated in the
$^2$H($^8$He,$^3$He)$^7$H reaction.
The features of the setups employed in individual experiments will be described
in subsections below.
The main contribution to the $^{7}$H MM energy resolution is the accuracy of the
energy determination of the $^3$He recoil, mainly caused by the uncertainty of
the interaction point $Z$-coordinate in the target volume.
For the purposes of the data analysis, it was assumed that the interaction point
was in the middle plane of the target.
To ensure a homogeneous thickness of the target, only events when the RIB hit a
central part of the target with a circular shape of the diameter of 17 mm were
taken into account.
This selection ensured also the rejection of the reactions with the material of
the target frame.


\subsection{Experiment 1}
\label{sec:exp-1}

\begin{figure}
\begin{center}
\includegraphics[width=0.47\textwidth]{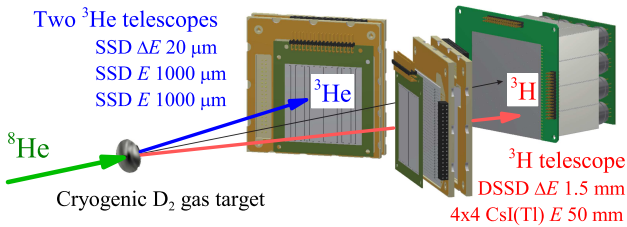}
\end{center}
\caption{Charged particle detector telescopes used in the experiment 1.}
\label{fig:setup-1}
\end{figure}

Two identical $\Delta E$-$E$-$E$ single-sided silicon-detector telescopes
provided the measurement of the $^3$He recoil nuclei emitted from the target
between $8^{\circ}$ and $26^{\circ}$ in the laboratory system, see Fig.\
\ref{fig:setup-1}.
Both telescopes located 166 mm downstream the target consisted of three layers
of silicon strip detectors (SSD).
The 20-micron-thick SSD with a sensitive area of $50 \times 50$ mm$^2$ was
divided into 16 strips, the second and the third layers were created by the two
identical 1-mm thick SSDs ($60 \times 60$ mm$^2$ with 16 strips).

The central telescope was installed at the beam line at the distance of 280\,mm
behind the target.
It was intended to detect tritons emitted with high energies at the angles
$\leq9^{\circ}$ in the laboratory system as a result of the $^7$H decay.
The telescope consisted of one 1.5\,mm thick double-sided SSD ($64 \times 64$
mm$^2$, with 32 strips on each side) followed by a square array of 16 CsI(Tl)
crystals.
The crystals had a cross section of $16.5\times16.5$ mm$^2$ and thickness 50\,mm
each, which allowed to stop all charged particles in the sensitive volume of the
telescope.
Each crystal was covered with a 3.5\,$\mu$m-thick aluminized Mylar on its
entrance and was coupled with its Hamamatsu R9880U-20 photomultiplier tube (PMT)
by the optical grease.
In order to increase the collection of light and to avoid light cross-talks,
each crystal was wrapped in a 100\,$\mu$m-thick VM-2000 reflector.

The $^{7}$H MM spectrum was reconstructed from measured energies and angles of
the $^{3}$He recoil particles detected in coincidence with the $^{3}$H fragments
\cite{Bezbakh:2020}.
The $^7$H MM spectrum was obtained from 119 $^{3}$H-$^{3}$He coincidence events.
The 1.1\,MeV energy resolution of the $^7$H MM spectrum \cite{Bezbakh:2020} was
much better as compared to the previous works
\cite{Korsheninnikov:2003,Caamano:2007,Caamano:2008, Nikolskii:2010}.


\subsection{Experiment 2}
\label{sec:exp-2}

\begin{figure}
\begin{center}
\includegraphics[width=0.47\textwidth]{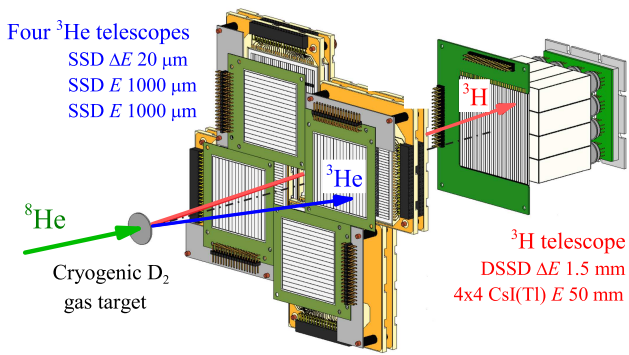}
\end{center}
\caption{The charged particle detector telescopes used in the experiment 2.}
\label{fig:setup-2}
\end{figure}

The most important task of the new experiment, studying the same
$^2$H($^8$He,$^3$He)$^7$H reaction, was to increase the statistics obtained in
the experiment 1 and to expand the measured angular range of the $^{3}$He
recoils to lower values in the laboratory system.
For this purpose, the detector setup was modified, see Fig.\ \ref{fig:setup-2}.
The new  $^{3}$He telescope assembly was installed at a distance of 179\,mm from
the target.
It consisted of four identical $\Delta E$-$E$-$E$ telescopes made of the same
SSDs as described in Section \ref{sec:exp-1}.
The angular range covered by these telescopes for the $^3$He recoil nuclei was
extended up to the range from $\sim 6^{\circ}$ to $\sim 24^{\circ}$ in the
laboratory system.

The tritons originating from $^{7}$H were emitted in the experiment 2 within
more narrow cone in comparison with the experiment 1.
Therefore, the central telescope, the same as in the experiment 1, was placed at
a distance of 323\,mm from the target.
As a result of these modifications, we could expect that the $^{7}$H yield
($^{3}$H-$^{3}$He coincidences) is increased by a factor of $\sim 2.5$.
Based on the results of the first experiment \cite{Bezbakh:2020} and taking into
account the expected larger cross section for the $^7$H population at smaller
angles, we expected that $\sim 300$ of $^{7}$H events should be collected in the
experiment 2.
This estimate corresponds well to the actually collected statistics of $378$
events.

The setup of experiment 2 also included the neutron spectrometer, made of 48
organic scintillator modules \cite{Bezbakh:2018}.
The spectrometer detects neutrons by measuring the light produced by the
interaction of the recoil charged particles (mainly protons) within the
scintillator.
It was located at zero angle in approximately 2 meters behind the reaction
chamber.
The distance between the neighboring modules was approximately 12\,cm, which
allowed to cover most of the forward angles, see Fig.\ \ref{fig:exp-common}.
The sensitive part of each module was cylinder made of stilbene monocrystal,
C$_{14}$H$_{12}$.
Each cylinder had 8\,cm diameter and 5\,cm thickness and was oriented by its
axis to the target.
Each crystal, covered with reflective MgO powder, was inserted into the 0.5\,mm
thick aluminum housing and connected to the PMT by the glass window and optical
grease.
Two types of PMT were used: Philips Photonics XP 4312 and ET-Enterprise 9822B.
In order to decrease the background signals produced by charged particles or
$\gamma$-rays, PMT-crystal systems were put into the steel tubes with 0.5\,mm
entrance windows.


\subsection{Reliability of channel identification and background conditions}
\label{sec:exp-2-channel}

\begin{figure}
\begin{center}
\includegraphics[width=0.49\textwidth]{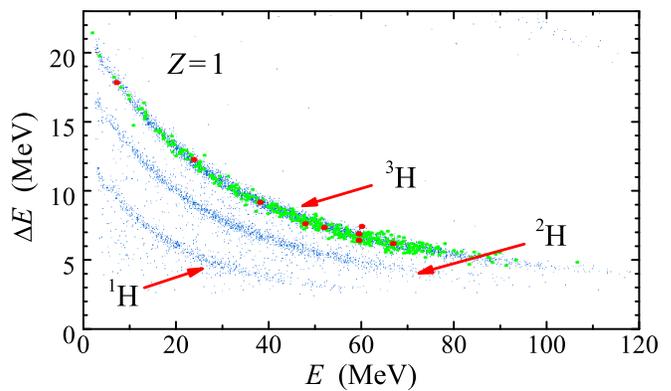}
\end{center}
\caption{
Identification of $^{3}$H by $\Delta E$-$E$ method in the central telescope.
The green dots show the double-coincidence $^{3}$He-$^{3}$H events, assigned to
the $^{7}$H spectrum in Fig.\ \ref{fig:exp-2-all} (b).
The large red dots indicate $^{7}$H g.s.\ candidate events, see discussion of
Section \ref{sec:exp-dis-2}.
}
\label{fig:exp-deltaee-3h}
\end{figure}
\begin{figure}
\begin{center}
\includegraphics[width=0.49\textwidth]{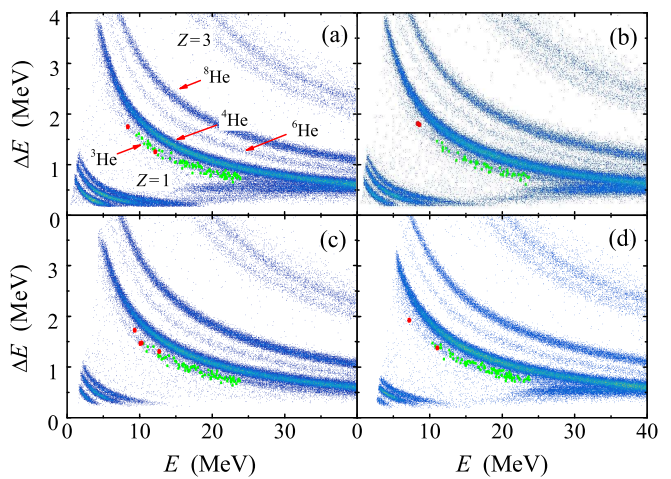}
\end{center}
\caption{
Identification of $^{3}$He recoil nuclei by $\Delta E$-$E$ method in four side
telescopes.
The green dots show the double-coincidence $^{3}$He-$^{3}$H events, assigned to
the $^{7}$H spectrum in Fig.\ \ref{fig:exp-2-all} (b).
The large red dots indicate the $^{7}$H g.s.\ candidate events, see discussion
of Section \ref{sec:exp-dis-2}.
}
\label{fig:exp-deltaee-3he}
\end{figure}

The background reduction  and unambiguous reaction channel identification were
the primary objectives of the experiment 2 because of the low statistics
obtained in the experiment 1 (5 ground state candidate events for the $1.8$\,MeV
state and $\sim 25$ events for the 6.5\,MeV state).
The most of the discussions of this work are based on the double-coincidence
$^{3}$He-$^{3}$H events.
The quality of $^{3}$H and $^{3}$He identification is illustrated in Figs.\
\ref{fig:exp-deltaee-3h} and \ref{fig:exp-deltaee-3he}, respectively.

Operation of the central $^{3}$H telescope was fairly standard, and the provided
particle identification is extremely reliable.
In contrast, a sophisticated analysis procedure was developed for the side
$^{3}$He telescopes.
A very thin 20\,$\mu$m $\Delta E$ detector is needed for the  $^{3}$He
identification, since the energy of the recoils leaving the target volume in the
laboratory system was expected to be very low (starts from $\sim 7$\,MeV).
Because of fabrication inhomogeneity, inherent for such a thin silicon plates,
the calibration thickness maps were determined for each of these thin detectors
\cite{Muzalevskii:2020}.
Fig.\ \ref{fig:exp-deltaee-3he} illustrates the particle identification, which
was actually implemented for each strip separately, but even in this
presentation it looks quite convincing.

Extremely strong background cleaning and channel identification for the $^{7}$H
population is provided by additional coincidences with neutrons. Statistics of
these measurements is extremely low, and they can be used in Fig.\
\ref{fig:exp-2-all} (c) just to demonstrate the compatibility of these data with
a suggested interpretation.

\begin{figure}
\begin{center}
\includegraphics[width=0.43\textwidth]{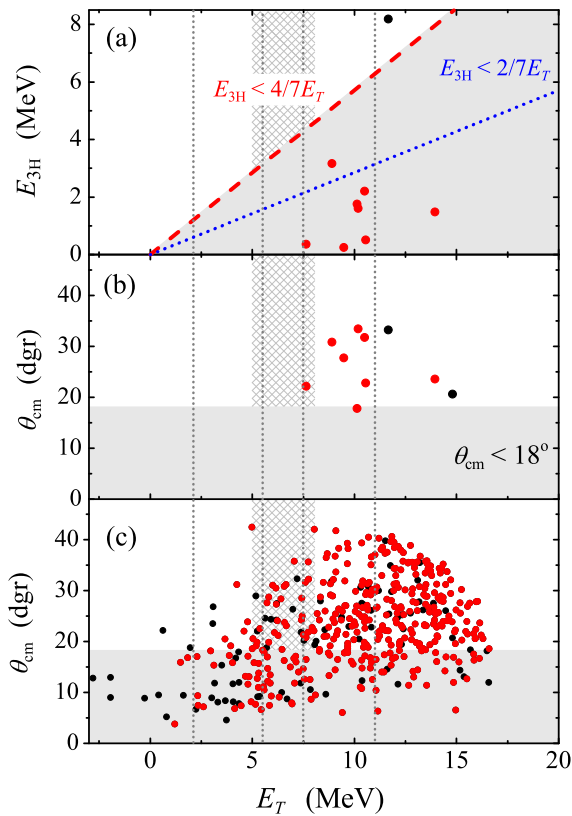}
\end{center}
\caption{
Empty target correlated spectra  $E_{3\text{H}}$ vs.\ $E_T$ for $^{7}$H (a) and
$\theta_{\text{cm}}$ vs.\ $E_T$ (b) for $^{7}$H in the experiment 2, resonant
states, see Fig.\ \ref{fig:exp-2-all}.
Panel (c) shows all the data of experiment 2 in the plane $\theta_{\text{cm}}$
vs.\ $E_T$.
In all panels the red dots show the distribution for the events within the
``kinematical triangle'', see Fig.\ \ref{fig:exp-2-all}; additional black dots
show the rest of the data.
The vertical gray dotted lines indicate assumed positions of the $^{7}$H
resonant states.
The vertical hatched area contains events either from ``asymmetric'' 5.5\,MeV
state or from the 5.5 -- 7.5\,MeV doublet.
The line $E_{\text{3H}}<2/7E_T$ in  panel (a) is discussed in Section
\ref{sec:exp-t-endis}.
}
\label{fig:exp-empty}
\end{figure}

Measurements with the empty target are standard approach to demonstrate directly
the background conditions of the experiment, see Figs.\ \ref{fig:exp-empty}
(a,b) and \ref{fig:exp-2-all} (b).
The measured empty target beam-integral values made $\sim 10 \%$ and $\sim 15
\%$ of the total beam-integral in the experiments 1 and 2, respectively.
In the experiment 1 only 3 empty target events were recorded, which allows to
evaluate the total background contribution in this experiment as $\sim 15 \%$.
In the experiment 2 this type of background is evaluated to be considerably
smaller, $\sim 8 \%$.
In the distribution of empty target events, see Fig.\ \ref{fig:exp-empty} (a), a
region within $7.5<E_T<10.5$\,MeV can be spotted.
Fortunately, these events are all concentrated on a certain angular range
$18^{\circ}< \theta_{\text{cm}}< 35^{\circ}$, see Fig.\ \ref{fig:exp-empty} (b).
Concentration of empty target events in the narrow region can hardly be
explained by statistical fluctuations, see the distribution of the complete data
in Fig.\ \ref{fig:exp-empty} (c).
This situation motivates us to avoid this region in the interpretation of the
data and enhance the confidence in the rest of the data.

Here we should remind briefly the situation with the reaction-channel
identification and background conditions in the previous experiments on proton
removal from $^{8}$He.
In experiment \cite{Korsheninnikov:2003} only the MM spectrum of $^{7}$H was
available, and background conditions were very poor: the MM spectrum extended
into negative energy region down to $-20$\,MeV, and more than $90 \%$ of the
data were related to the background in the analysis.
In experiment \cite{Fortier:2007} also only the MM spectrum of $^{7}$H was
available, and background conditions were poor: the MM spectrum extended into
the negative energy region down to $-3$\,MeV, and $\sim 75 \%$ of the data were
related to the background reactions originated from the carbon component of the
CD$_2$ target.
In the experiment \cite{Nikolskii:2010} the MM spectrum of $^{7}$H was augmented
by the requirement of the $^{3}$He-$^{3}$H  coincidence, which drastically
improved the background conditions.
Still, some evidence of the background is visible, since the MM spectrum extends
into the negative region beyond the values implied by the energy resolution of
the experiment.
In the experiments presented in this work, the coincidence with $^{3}$H and the
reconstruction of the $^{3}$H momentum allow for using the kinematical triangle
condition as a selection gate, which reduces significantly the MM background,
see Fig.\ \ref{fig:exp-2-all} (a,d) and Section \ref{sec:exp-t-endis}.


\subsection{Experimental resolution}
\label{sec:exp-resol}

\begin{figure}
\begin{center}
\includegraphics[width=0.47\textwidth]{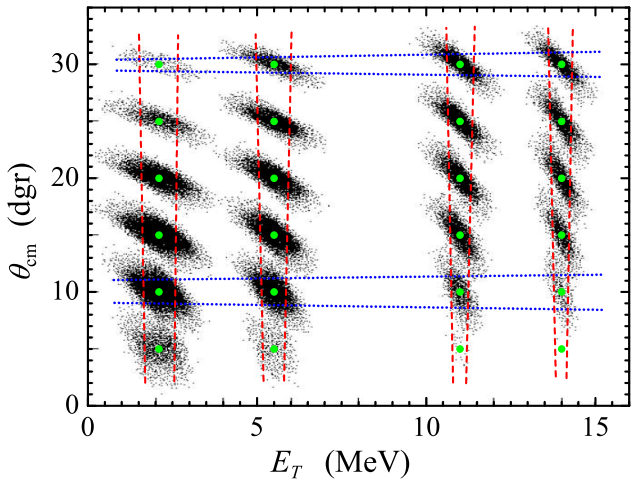}
\end{center}
\caption{
The Monte-Carlo simulations demonstrating the experiment 2 setup resolution for
the $^2$H($^8$He,$^3$He)$^{7}$H reaction.
The simulations are performed for a set of fixed center-of-mass reaction angles
$\theta_{\text{cm}}$ and the $^{7}$H decay energies $E_T$ indicated by green
dots.
The red dashed lines guide an eye along the contours defining the FWHM the for
energy resolution, while the blue dotted lines do the same for the angular
resolution.
}
\label{fig:delta-tcm-vs-e7h}
\end{figure}

The complete MC simulations of the experimental setup were performed and
extensively used in the interpretation of the data.
Here we address the question of experimental resolution.
The Fig.\ \ref{fig:delta-tcm-vs-e7h} shows MC simulations for the angular
$\theta_{\text{cm}}$ vs.\ energy $E_T$ distributions defined by the product of
$\delta$-functions at the corresponding energy and angle.
The projections of the plotted structures either on the energy or the angle axis
reflect the respective resolutions at a certain place of the kinematic plane,
see Table \ref{tab:delta-tcm-vs-e7h}.
It is possible to find out that at $\theta_{\text{cm}} \rightarrow 0$ the energy
resolution is defined mainly by the target thickness.
The relative importance of this factor decreases with increase of the $^{7}$H MM
energy: the energy resolution is changing from $\sim 800$ keV at $E_T=2.2$\,MeV
to $\sim 250$ keV at $E_T=14$\,MeV.
The angular resolution at $\theta_{\text{cm}} \rightarrow 0$ is defined by the
beam tracking precision and granularity of the $^{3}$He telescopes.
It is clear from Fig.\ \ref{fig:delta-tcm-vs-e7h}, that for large
$\theta_{\text{cm}}$ the MC spots are tilted and, thus, both the energy and the
angular resolutions aggregate the two mentioned factors.
Consequently, the best resolution for the $^{7}$H g.s.\ MM energy is obtained
for the small center-of-mass reaction angles, and for the larger angles it
considerably degrades.

\begin{table}[b]
\caption{
Experimental resolution in the second experiment as a function of the $^{7}$H MM
energy and center-of-mass angle $\theta_{\text{cm}}$ based on the MC simulations
Fig.\ \ref{fig:delta-tcm-vs-e7h}.
The first and second values in each cell are the FWHM energy and the angular
resolutions given in\,MeV and degrees, respectively.
}
\begin{ruledtabular}
\begin{tabular}[c]{ccccccccc}
$E_T$   & \multicolumn{2}{c}{2.2\,MeV} &   \multicolumn{2}{c}{5.5\,MeV} &
  \multicolumn{2}{c}{11\,MeV} &   \multicolumn{2}{c}{14\,MeV} \\
\hline
 $10^{\circ}$ & 0.95 & 2.2 & 0.73 & 2.3 & 0.48 & 2.5 & 0.38 & 2.8  \\
 $20^{\circ}$ & 1.10 & 1.6 & 0.93 & 1.8 & 0.64 & 2.2 & 0.52 & 2.6  \\
 $30^{\circ}$ & 1.13 & 1.2 & 0.99 & 1.3 & 0.77 & 1.8 & 0.69 & 2.0  \\
\end{tabular}
\end{ruledtabular}
\label{tab:delta-tcm-vs-e7h}
\end{table}


\subsection{Calibration $^2$H($^{10}$Be,$^3$He)$^{9}$Li
reaction}
\label{sec:exp-2-10be}

\begin{figure}
\begin{center}
\includegraphics[width=0.45\textwidth]{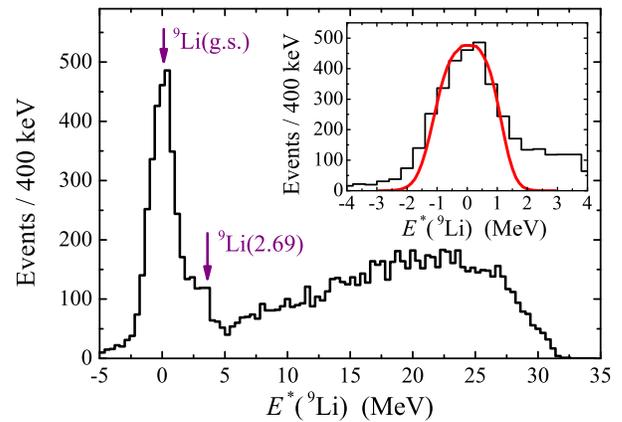}
\end{center}
\caption{
Excitation spectrum of $^{9}$Li  measured in the $^2$H($^{10}$Be,$^{3}$He)$^9$Li
reaction.
The insert shows the part of the $^{9}$Li spectrum near the ground state.
The red curve represents the Monte-Carlo calculation of the $^{9}$Li(g.s.)
taking the parameters of the experimental setup.
}
\label{fig:9li}
\end{figure}

The proton pick-up reaction ($^2$H,$^3$He) was studied with a 42\,AMeV $^{10}$Be
secondary beam in order to test the reliability of the obtained experimental
data on the $^{7}$H  system, to control calibration parameters, and to get an
experimental estimate of the $^{7}$H MM resolution.
These measurements were performed just after the experiment 2 and all conditions
related to the experimental setup (experiment 2) were kept the same.
The excitation spectrum of $^9$Li derived from the data of the $^{10}$Be run is
shown in Figure \ref{fig:9li}.

The solid histogram in Fig.\ \ref{fig:9li} shows a well-pronounced peak
corresponding to the g.s.\ of $^9$Li populated in the
$^2$H($^{10}$Be,$^{3}$He)$^9$Li(g.s.) reaction.
On the right slope of this peak the population of not well-resolved first
excited state of $^9$Li ($E^* = 2.69$\,MeV) is also observed.
The insert in Fig.\ \ref{fig:9li} shows the part of the $^9$Li spectrum near the
g.s.
The red curve demonstrates the Monte-Carlo calculation for the $^9$Li(g.s.)
using parameters of the experimental setup.

It may be clearly seen that the MC simulation reproduces quite well the shape of
the $^9$Li(g.s.) peak demonstrating the resolution of $\sim 2.2$\,MeV (FWHM).
The corresponding calculations of the MM resolution of $^{7}$H at energy near
2\,MeV gave FWHM $\sim 1.1$\,MeV (see discussion in Section III).
The reason for this $\sim 2$ times better resolution in the $^7$H experiment is
caused by the larger energies of $^3$He recoils, as compared to the
$^2$H($^{10}$Be,$^{3}$He)$^9$Li reaction, and, therefore, the smaller energy
losses in the target.
It is also a demonstration that the target thickness makes the main contribution
to the energy resolution in this energy range.
The cross-section values of $\sim 7 - 10$ mb/sr at forward angles in the c.m.\
system were deduced from these data for the reaction populating the $^9$Li
ground state.

Thus, the data obtained with the $^{10}$Be beam provide an independent
cross-check of the MM spectrum calibration in the experiment 2 and validation
for the developed MC simulation framework.
This is an important support of the data and interpretation of the experiment 2,
which was not available for the experiment 1.


\section{Discussion of the $^{7}$H data}
\label{sec:exp-dis}

\begin{figure}
\begin{center}
\includegraphics[width=0.48\textwidth]{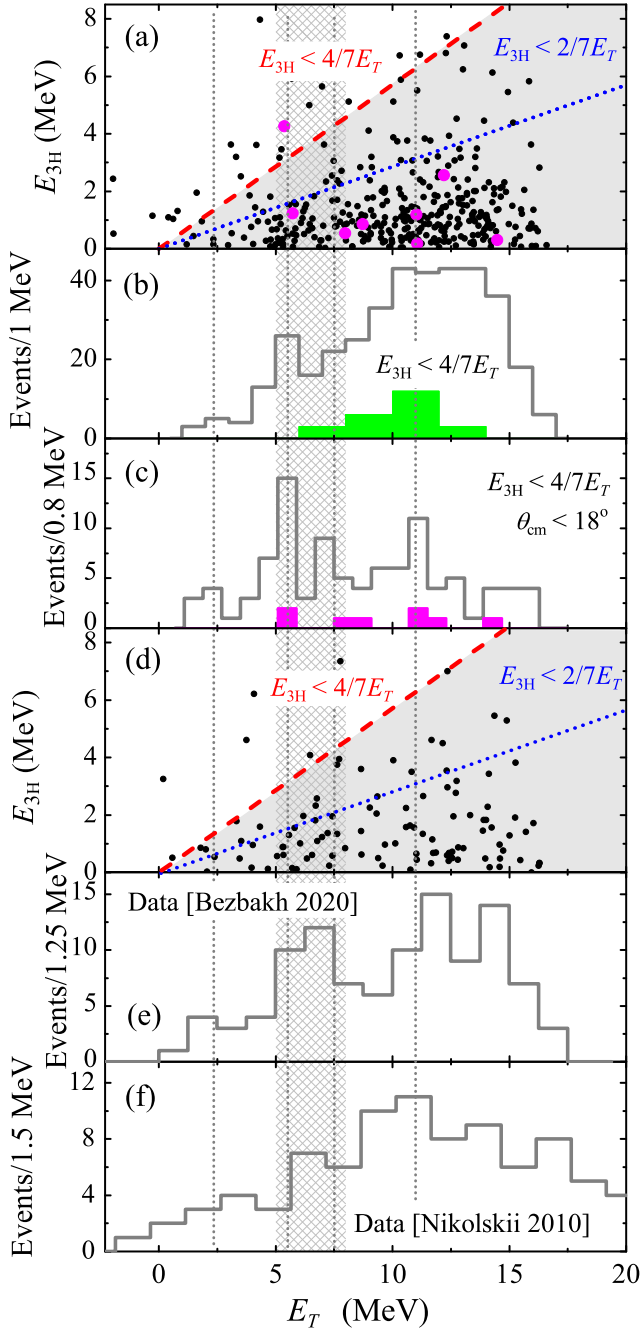}
\end{center}
\caption{
(a) Correlation between the $^3$H energy in the $^{7}$H frame and the $^{7}$H
decay energy derived from the data of experiment 2 (black circles).
The large magenta circles show triple coincidence $^{3}$He-$^{3}$H-$n$ events.
The shaded area corresponds to events matching a condition of maximal energy
$E_{3\text{H}} < 4/7E_T$ possible in the decay of $^{7}$H.
(b) The $^{7}$H MM spectrum projected from (a) by using a gate condition with
cutoff $E_{3\text{H}} < 4/7 E_T$.
The green-filled histogram shows the background inferred from the empty target
data Fig.\ \ref{fig:exp-empty} (a).
(c) The $^{7}$H MM spectrum projected  from (a) with the two selection gates,
$E_{3\text{H}} < 4/7 E_T$ and $\theta_{\text{cm}}< 18^{\circ}$.
The magenta-filled histogram indicates the triple coincidence
$^{3}$He-$^{3}$H-$n$ events.
(d) Correlation between the $^3$H energy in the $^{7}$H frame and the decay
energy of $^{7}$H from Ref.\ \cite{Bezbakh:2020}.
(e) The $^{7}$H MM-derived spectrum from Ref.\ \cite{Bezbakh:2020}.
(f) The $^{7}$H MM spectrum from Ref.\ \cite{Nikolskii:2010}.
The vertical gray dotted lines indicate the assumed positions of the $^{7}$H
resonant states.
The vertical hatched area contains events either from ``asymmetric'' 5.5\,MeV
state or from the 5.5 -- 7.5\,MeV doublet.
The line $E_{\text{3H}}<2/7E_T$ in panel (a) is discussed in Section
\ref{sec:exp-t-endis}.
}
\label{fig:exp-2-all}
\end{figure}

The survey of the $^{7}$H spectra obtained by using the MM method in this work,
as well as in Refs.\ \cite{Nikolskii:2010,Bezbakh:2020}, is given in Fig.\
\ref{fig:exp-2-all}.
From these spectra we assign the ground state at $2.2(3)$\,MeV, the first
excited state at 5.5(3)\,MeV, and the higher-energy resonances at 7.5(3) and
11.0(5)\,MeV.

At first glance, the resonance features in the $^{7}$H  MM spectrum in Fig.\
\ref{fig:exp-2-all} (b) are not very pronounced.
For that reason we provide first a general note about the observation of
resonant states in a spectrum either containing broad overlapping states or
having important continuous background contribution, and then turn to a detailed
inspection of our data.


\subsection{General note on resonant states observations}

\begin{figure*}
\begin{center}
\includegraphics[width=0.999\textwidth]{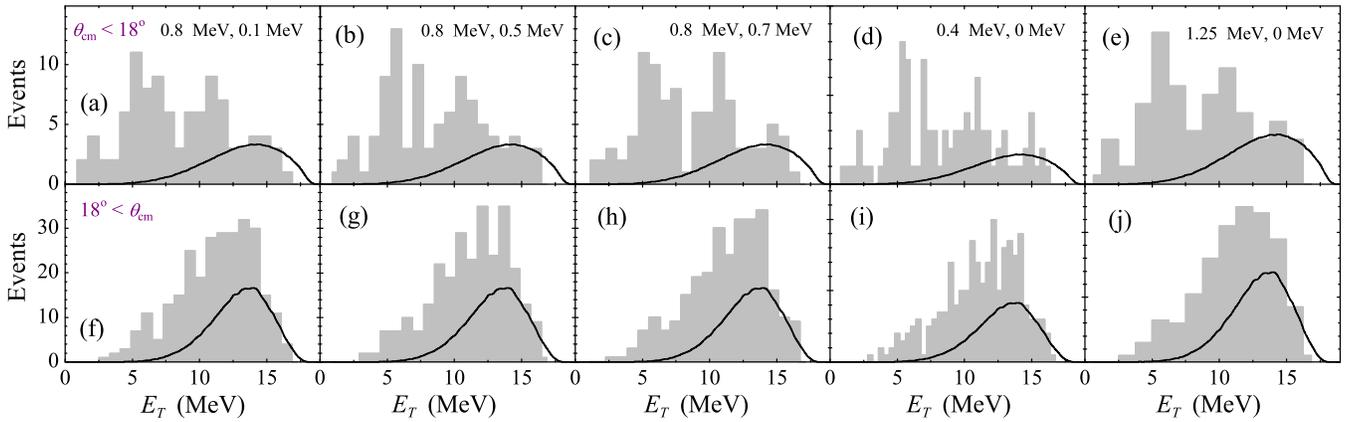}
\end{center}
\caption{
The missing mass spectrum obtained in the experiment 2, in different
representations.
Two c.m.\ angular ranges are selected: the top row shows events selected by the
gate $\theta_{\text{cm}} < 18^{\circ}$, the bottom row shows spectra selected by
the $\theta_{\text{cm}} > 18^{\circ}$ condition.
The binning factor and bin offset are shown for each column in the top row.
The ``phase-volume'' curves $d \sigma /dE_T d \Omega \sim E^5_T$ with
experimental bias accounted by MC procedure are shown in the both rows
(normalization is arbitrary).
}
\label{fig:rebin-test}
\end{figure*}

Let us consider one selected spectrum from Fig.\ \ref{fig:exp-2-all}.
Is it possible to interpret it without an assumption about the population of
resonant states?
To answer this question in the first approximation, different representations of
the $^{7}$H MM spectrum are shown in Fig.\ \ref{fig:rebin-test} with own binning
factors and bin offsets.
The $^{7}$H decay events with $\theta_{\text{cm}}<18^{\circ}$ were selected, see
Fig.\ \ref{fig:exp-2-all} (c), and we accordingly split the data in two parts in
Fig.\ \ref{fig:rebin-test}.
One motivation for $\theta_{\text{cm}}=18^{\circ}$ selection is illustrated in
Fig.\ \ref{fig:delta-tcm-vs-e7h}: the best energy resolution of the $^{7}$H
spectrum is obtained for the small center-of-mass reaction angles, and it
considerably deteriorates at larger angles.
The selected $\theta_{\text{cm}}<18^{\circ}$  range is also consistent with the
cutoff needed for elimination of a ``dangerous background region'', specified in
the experiment with empty target, see Fig.\ \ref{fig:exp-empty} (c).
For $\theta_{\text{cm}}<18^{\circ}$ the three resonant structures at 2.2, 5.5,
and 11\,MeV are well identified in all representations in Fig.\
\ref{fig:rebin-test}.
Evidence of the 7.5\,MeV peak may be statistically insignificant in some
representations, but it is typically present.
So, the assumed resonant structures are at least not artefacts of the histogram
arrangement.
The spectra with the  $\theta_{\text{cm}}>18^{\circ}$ selection gate are
dominated by a smooth ``phase-volume''-like contribution.
Only the 5.5\,MeV peak can be clearly seen on the top of the smooth component.
Some resonance contributions can be suspected at energies $E_T > 10$\,MeV, but
their manifestation on the  top of the large smooth component is statistically
insignificant.

Then we turn to statistical analysis.
It shows that the description of each spectrum in Fig.\ \ref{fig:exp-2-all}
merely by some smooth underlying continuum is possible with values of
root-mean-square deviation (RMSD) for the spectrum in Fig.\ \ref{fig:exp-2-all}
(f) with RMSD $ \sim 1$,  for the spectrum in Fig.\ \ref{fig:exp-2-all} (e) with
RMSD $ \sim 1-2$, and for the spectra in Figs.\ \ref{fig:exp-2-all} (b,c) with
RMSD $\sim 2-3$.
These are statistically tolerable values of the mismatch, which does not exclude
a ``smooth scenario''.
However, the following general points should be clarified.

If the real spectrum of $^{7}$H is smooth, then, due to the small-statistics
data, a purely random mockup of several peaks is possible.
In such a case data with very large statistics (e.g.\ $10^3-10^4$ events) are
required in order to exclude such accidental ``resonances'' with a high
confidence level.
In contrast, if the real $^{7}$H spectrum contains narrow resonant peaks, then
reliable identification of these resonant states becomes possible even with few
measured decay events.
We assume that the $1/2^+$ g.s.\ of $^{7}$H and the lowest excitations, such as
$5/2^+$--$3/2^+$ doublet are located at $E_T < 10 $\,MeV.
The width estimates for such $^{7}$H states provided in the related article
\cite{Grigorenko:2021} show that the widths are likely to be quite small with
the expected values of $\Gamma \lesssim 1$\,MeV.
So, the narrow resonant-state scenario seems to be physically reasonable and
even unavoidable at least for $E_T < 10$\,MeV.

One can see in Fig.\ \ref{fig:exp-2-all} that the same peaks may be spotted in
all three experimental datasets of the $^2$H($^8$He,$^3$He)$^{7}$H reaction.
The individual statistics of the datasets of the order $100-400$ events can not
exclude a pure statistical origin of these peaks in each case.
However, it is extremely improbable that the same statistical artefacts could
arise in the three different, totally independent experiments.
This is a strong general argument supporting the data interpretation of this
work.
Below we provide an in-depth view in different aspects inherent to each
structure and also demonstrate that all these aspects can be interpreted in a
consistent way.


\subsection{Group of events at 2.2\,MeV}
\label{sec:exp-dis-2}


The events with $E_T<3.2 $\,MeV were selected as candidates to represent the
$^{7}$H ground state.
There are 9 such events with the mean energy value of $2.2$\,MeV and the
dispersion of $0.6$\,MeV.
These values agree well with the results of the experiment 1 reported in Ref.\
\cite{Bezbakh:2020}, where the $^{7}$H g.s.\ energy $E_T=1.8(5)$\,MeV was
obtained.
The events are well separated (there is $\sim 0.5$\,MeV gap) from the nearest
event with the higher $^{7}$H excitation.
There are 4 possible reasons to get these events here:  (i) background events,
(ii) ``contamination'' by events from the higher excitations of $^{7}$H, (iii)
some smooth ``phase-volume-like'' distribution, (iv) narrow resonant state.
We accept the option (iv), but we have to comment on the other points as well.

\noindent (i) Possible background contribution in the $E_T$ region of interest
can be estimated on the basis of the empty target measurements.
No background events were observed in proximity, see Fig.\ \ref{fig:exp-empty}
(a,b).
Another option is to estimate it from the density of background counts beyond
the kinematical triangle in Fig.\ \ref{fig:exp-2-all} (a).
Here we can expect $ \sim 1$ background event in the 2.2\,MeV group.

\noindent (ii) The observed width of the 2.2\,MeV event group is assumed to be
entirely defined by the energy resolution of the experiment.
The theory estimates \cite{Grigorenko:2011,Grigorenko:2021} give $\Gamma
\lesssim 1$ keV  for the $^7$H g.s., so this assumption is true even if the
actual intrinsic width of this state is larger by a factor $100 - 500$, as
compared to the existing estimates.
The discussion of the energy profiles of the $^{7}$H first excited state is
provided in the Section \ref{sec:exp-dis-6} and in Ref.\ \cite{Grigorenko:2021}.
The given there theoretical estimates agree that there should be an empty
``window'' between the ground state and the first excited state of the $^{7}$H
extending from $E_T \sim 3$\,MeV to $E_T \sim 4.0-4.5$\,MeV.
Any events emerging in this energy range should be connected with some
background or/and the MM resolution.
The MC simulations of the $^{7}$H MM spectrum are shown in Fig.\
\ref{fig:profile-for-simulations}.
They confirm that even the weakly populated $^{7}$H g.s.\ can be reliably
separated from the ``tail'' of the first excited state, and that such a
separation is the best at the small c.m.\ angles of the
$^2$H($^8$He,$^3$He)$^7$H reaction.

\noindent (iii) The phase-volume-like energy dependence of cross sections can be
expected if the spectral properties of the coresponding continuum are not
expressed at all.
The phase-volume for the true core+$4n$ decay at $\sim 2.2$\,MeV is
\begin{equation}
dW/d E_T \sim  E_T^7 \,.
\label{eq:e7}
\end{equation}
The ordinarily expected five-body phase-volume is $\sim E^5$, but the required
in our specific case, four-neutron anti-symmetrization modifies it to formula
(\ref{eq:e7}), see Ref.\ \cite{Grigorenko:2011}.
At $E_T \sim 2.5 - 3$\,MeV a turnover may occur from the true $4n$ emission to
the sequential two-neutron emission via the $^{5}$H ground state (located at
about 1.8\,MeV above the $^{3}$H+$n$+$n$ breakup threshold
\cite{Korsheninnikov:2001,Golovkov:2005}).
So, at some higher energy the dependence
\begin{equation}
dW/d E_T  \sim  E_T^2 \,,
\label{eq:pv-0p}
\end{equation}
characteristic for the three-body decays, may take place.
In any case, the phase-volume behavior that can be expected for $^{7}$H (say, at
$E_T<10 $ MeV) is a strongly growing function of energy, starting straight from
the core+$4n$ threshold.

An additional support for the interpretation of the group of events at
$2.2$\,MeV as a resonant state is provided by the three types of distributions
which are analyzed for the events of the 2.2\,MeV group in the following
Subsections: (i) the $^{7}$H center-of-mass angular distribution, obtained for
the $^2\text{H}(^8\text{He},^3\text{He})^{7}$H reaction, (ii) $^{3}$H energy
distribution in the $^{7}$H frame, and (iii) $^{3}$H lab system angular
distribution relative to the $^{7}$H momentum direction.
Statistics, which we have for the ground-state candidate events, is very small.
However, all the mentioned distributions demonstrate correlated character,
expected for the $^{7}$H g.s.\ decay, in contrast with the casual distribution
expected for the background events.

\subsubsection{$^{7}$H center-of-mass angular distribution}
\label{sec:exp-angdis-gs}

The center-of-mass angular distributions for the $^2$H($^8$He,$^3$He)$^{7}$H
reaction are further discussed in Section \ref{sec:exp-angdis}.
Here we address the distribution of the 2.2\,MeV event group.

\begin{figure*}
\begin{center}
\includegraphics[width=0.99\textwidth]{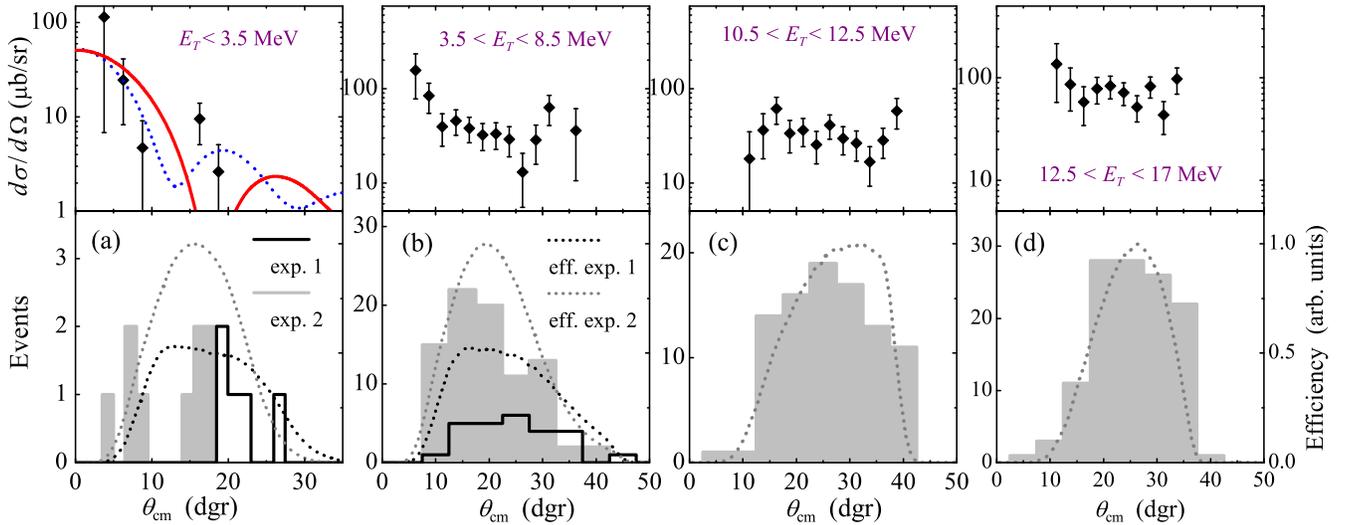}
\end{center}
\caption{
The center-of-mass angular distributions for the
$^2\text{H}(^8\text{He},^3\text{He})^{7}$H reaction in different $^{7}$H MM
energy ranges.
Efficiency corrected angular distributions are shown in the upper part of each
panel.
(a) The $^{7}$H g.s.\ with $0 < E_T< 3.2$\,MeV and $E_{3\text{H}}<4/7E_T$, the
hollow black and filled gray histograms shows the data of the experiments 1 and
2, respectively.
The dotted curves of corresponding color show the detection efficiency which is
given on the right axis.
The red solid curve shows the FRESCO calculation results with standard
parameters, the blue dashed curve shows the result of FRESCO calculation
assuming the extreme peripheral transfer (both curves have arbitrary scaling).
(b) The first excited state (or doublet) with $3.2 < E_T< 8.5$\,MeV.
(c) The 11\,MeV excited state with $10.5 < E_T< 12.5$\,MeV.
(d) The high-energy part of the spectrum with $E_T > 12.5$\,MeV.
}
\label{fig:angdis}
\end{figure*}

In the $^{7}$H g.s.\ case, there are no events at $\theta_{\text{cm}}$ from
$9.5^{\circ}$ to $15.5^{\circ}$, see Fig.\ \ref{fig:angdis} (a).
This feature is consistent with the observations of Ref.\ \cite{Bezbakh:2020},
where 5 g.s.\ candidate events were localized in the range
$18^{\circ}<\theta_{\text{cm}}<27^{\circ}$.
Such observed angular distributions can be problematic from the theoretical
point of view because theory typically predicts the diffraction minimum at
$\theta_{\text{cm}} \sim 16^{\circ}-18^{\circ}$ for the similar transfer
reactions.
The considerably lower cross-section-minimum position, at $\theta_{\text{cm}}
\sim 13^{\circ}-14^{\circ}$, is suggested by the data.
If the latter is true, then the observed angular distribution provides important
tip for the following problems.

\noindent (i) There was a problem pointed out in Ref.\ \cite{Bezbakh:2020}: the
g.s.\ angular acceptance of experiment 1 was high enough for $\theta_{\text{cm}}
> 8^{\circ}$ [see the black dotted curve in Fig.\ \ref{fig:angdis} (a)] to
ensure
the registration of few events in the angular range $\theta_{\text{cm}} \sim
8^{\circ}-15^{\circ}$ (assuming the diffraction minimum at $\theta_{\text{cm}}
\sim 16^{\circ}-18^{\circ}$).
However, no events were observed in this range.
One may explain this fact, if the diffraction minimum actually covers a range of
$10^{\circ}<\theta_{\text{cm}}<16^{\circ}$.

\noindent (ii) The DWBA/FRESCO calculations with more or less standard
parameters fail to provide the diffraction minimum at $\theta_{\text{cm}} \sim
13^{\circ}-14^{\circ}$.
Such a small angle of the diffraction minimum might provide an evidence for the
extreme peripheral character of the $^{7}$H g.s.\ population in the
$^2$H($^8$He,$^3$He)$^{7}$H reaction.
The extreme peripheral character of the reaction, in turn, gives a natural
explanation of the extremely low cross section observed for the $^{7}$H g.s.\
population ($\sim 25$ $\mu$b/sr within the angular range $\theta_{\text{cm}}
\sim 17^{\circ}-27^{\circ}$ \cite{Bezbakh:2020}).
See also Ref.\ \cite{Grigorenko:2021} for an extended discussion of this point.

How statistically significant is the $\theta_{\text{cm}} \sim 9.5^{\circ} -
15.5^{\circ}$ gap in the g.s.\ angular distribution? Let's make a simple
estimate: by assuming that the actual angular distribution is homogeneous, and
the experimental efficiency is constant and nonzero in the ranges
$\theta_{\text{cm}} \sim 8^{\circ} - 26^{\circ}$ and $\theta_{\text{cm}} \sim
6^{\circ} - 24^{\circ}$ in the experiments 1 and 2, respectively.
Then the estimated probability of non-population of the $\theta_{\text{cm}} \sim
9.5^{\circ} - 15.5^{\circ}$ range in both experiments simultaneously is $\sim 1
\%$.
This estimate is likely to be an upper estimate of the corresponding
probability, since the setup efficiency is either large or close to maximal in
the $\theta_{\text{cm}} \sim 9.5^{\circ} - 15.5^{\circ}$, see Fig.\
\ref{fig:angdis} (a).
Thus the careful treatment of the experimental bias can only further decrease
this estimate.
So, it is very unlikely that the experimentally observed patterns are generated
by some featureless distribution occurring due to statistical fluctuations.
The interpretation by assigning the diffraction minimum at $\theta_{\text{cm}}
\sim 13^{\circ}-14^{\circ}$ is, thus, quite natural.

The best energy resolution of the $^{7}$H g.s.\ can be expected for the
small-angle events from the first diffraction maximum.
Indeed, by selecting 4 events with $\theta_{\text{cm}} < 10^{\circ}$ we obtain a
bit different mean energy $E_T=2.1$\,MeV and dispersion of 0.55\,MeV (compared
to results with the complete data set).
The dispersions of the g.s.\ events for the small $\theta_{\text{cm}}$ (0.55
MeV) and for the complete data (0.6 MeV) are consistent with the MC estimated
energy resolutions, see Table \ref{tab:delta-tcm-vs-e7h}.

\subsubsection{$^{3}$H energy distribution in the $^{7}$H rest frame}
\label{sec:exp-t-endis}

\begin{figure}
\begin{center}
\includegraphics[width=0.42\textwidth]{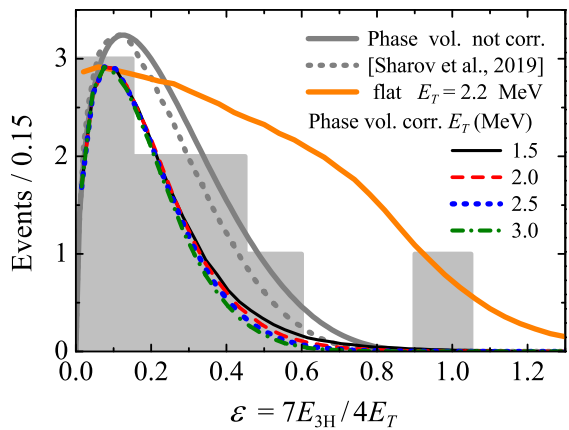}
\end{center}
\caption{
The $^{7}$H center-of-mass $\varepsilon$ distribution of the $^{3}$H fragments
emitted at the $^7$H$ \rightarrow$ $^{3}$H+4n decay.
The filled gray histogram is the the distribution obtained in the experiment 2
for the detected events with $E_T<3.2$\,MeV.
The gray solid and dotted curves show the theoretical (not corrected for
experimental response) phase-volume distribution [see Eq.\ (\ref{eq:pv-5b-7}]
and the realistic distribution from \cite{Sharov:2019}, respectively.
The curves labeled with the $^{7}$H  g.s.\ decay energy $E_T$ show the
MC-simulation results obtained for the phase-volume distribution of Eq. (3) at
different decay energy values supposed for the $^{7}$H g.s.\ resonance.
}
\label{fig:eps-dis}
\end{figure}

The emission dynamics of the true $4n$ nuclear decay is still a completely
unexplored phenomenon.
Our data for the first time provide access to this type of information.
The commonly expected energy distribution of the $^{3}$H fragments emitted at
the $^{7}$H g.s.\ decay, has the shape of a 5-body phase-volume
\begin{equation}
\frac{dW}{d \varepsilon} = \sqrt{\varepsilon (1-\varepsilon)^7} \,, \quad
\varepsilon = \frac{7E_{3\text{H}}}{4E_T} \,,
\label{eq:pv-5b-7}
\end{equation}
where $E_{3\text{H}}$ is the energy of $^{3}$H in the $^{7}$H rest frame.
This distribution suggests that $\sim 92 \%$ of events are located below
$\varepsilon=1/2$ and the mean $^{3}$H energy value $\langle \varepsilon \rangle
= 1/4$.
Moreover, a realistic energy distribution obtained in the 5-body calculations of
Ref.\ \cite{Sharov:2019} has even more correlated character, with $\langle
\varepsilon \rangle \sim 0.21-0.22$ for $E_T \sim 2-3$\,MeV, see Fig.\
\ref{fig:eps-dis}.
This happens because in the decay via the 4 neutrons emission, at least 2
additional excitation quanta in the ``neutron part'' of the  WF are needed to
enable the anti-symmetrization of the WF.
In this distribution $\sim 95 \%$ of events are located below $\varepsilon=1/2$.

The presented phase-volume argument tells us that the absolute majority of
``physical'' events, produced by the $^{7}$H g.s.\ decay, should, most likely,
reside below the ``$E_{\text{3H}} = 2/7 E_T$ line'' in Figs.\
\ref{fig:exp-empty} (a) and \ref{fig:exp-2-all} (a,d).
Thus, the events with $\varepsilon > 0.5-0.6$ are likely to be associated either
with some sort of background or with the poor resolution obtained for the
reconstructed energy of the $^{3}$H in the center-of-mass of the $^{7}$H.

The observed $^{3}$H distribution for the expected $^{7}$H g.s.\ events is shown
in Fig.\ \ref{fig:eps-dis}.
In this figure we compare the experimental data with the phase-volume
distributions derived at different energies $E_T$.
The gray solid curve shows the five-body phase-volume distribution of Eq.\
(\ref{eq:pv-5b-7}), which does not depend on energy $E_T$.
However, with the experimental bias taken into account via the MC procedure, the
``observable'' phase-volume distributions become somewhat different (black, red,
blue, and green curves) and energy dependent.
The result of the MC simulation for the uncorrelated ``flat'' energy
distribution is given in Fig.\ \ref{fig:eps-dis} by the solid orange curve.
The respective numerical information is also given in Table \ref{tab:3h-distr}.
One can find that the experimental energy distribution is consistent with the
expected for the $^{7}$H g.s.\ correlated emission and totally inconsistent with
the uncorrelated situation.
A more detailed discussion of the $^{3}$H energy distributions is provided in
Ref.\ \cite{Grigorenko:2021}.

\subsubsection{$^{3}$H-$^{7}$H angular distribution in the lab frame}
\label{sec:exp-t-andis}

From the theoretical point of view, such a distribution is directly connected
with the energy distribution of $^{3}$H in the $^{7}$H frame discussed in the
previous Section.
Moreover, the  $^{3}$H-$^{7}$H angular distribution is obtained by projecting
the $^{3}$H momentum on the transversal plane, so the information available in
the $^{3}$H energy distribution is partly lost here.
However, from the experimental point of view, this distribution is derived in a
methodologically different and more safe way because the reconstruction of the
$^{3}$H energy is not needed, only the $^{3}$H tracking should be done.
The MC evaluated resolution of this angular distribution is quite good $\Delta
\theta_{\text{3H-7H}} \lesssim 0.4^{\circ}$.

\begin{figure}
\begin{center}
\includegraphics[width=0.48\textwidth]{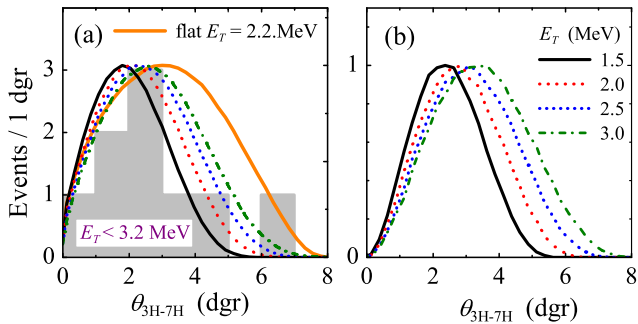}
\end{center}
\caption{
The angular distributions of the $^{3}$H fragments in the lab frame relative to
the $^{7}$H center-of-mass momentum vector.
The $^{7}$H g.s.\ candidate events with $E_T<3.2$\,MeV are selected.
(a) The experimental data and theoretical predictions with the experimental bias
taken into account by the MC simulation.
(b) The initial theoretical distributions based on phase-volume for different
$E_T$ values.
The curve designations are the same as in Fig.\ \ref{fig:eps-dis}.
}
\label{fig:ang-dis-3h}
\end{figure}

The angular distribution of $^{3}$H in the lab frame relative to the $^{7}$H
tracking obtained in the experiment 2 is shown in Fig.\ \ref{fig:ang-dis-3h}
together with the different predictions.
From this figure one may conclude that the experimental distribution is hardly
consistent with the smallest considered energy $E_T=1.5$\,MeV and is clearly
inconsistent with an uncorrelated distribution (``flat'' curve).
Finally, from information given in Table \ref{tab:3h-distr} one can find that
both, the energy and angular distributions of $^{3}$H, are consistent with the
$E_T=2.2(5)$\,MeV energy inferred from the MM data.

\begin{table}[b]
\caption{
Mean values of the $\varepsilon$ and $\theta_{\text{3H-7H}}$ variables for the
distributions in Figs.\ \ref{fig:eps-dis} and \ref{fig:ang-dis-3h} with varied
$E_T$ in the top row.
It is problematic to estimate errors of the experimental mean values.
However, error estimates can be done for the theoretical distributions based on
the MC simulations with the same statistics (9 events) as in the experiment.
The value $\langle \varepsilon \rangle$, obtained in the experiment, is
consistent with $E_T<2.2$\,MeV.
The best fit to the experimental $\langle \theta_{\text{3H-7H}} \rangle$ value
is obtained at $E_T =2.6(7)$\,MeV.
Both values are consistent with $E_T =2.2(5)$\,MeV inferred from the MM data.
}
\begin{ruledtabular}
\begin{tabular}[c]{ccccccc}
 Value         & flat & 1.5 & 2.0 & 2.5 & 3.0 & Exp. \\
\hline
 $\langle \varepsilon \rangle$  & 0.46(11) & 0.28(6) & 0.26(6) & 0.24(6) &
 0.23(6) & 0.31 \\
 $\langle \theta_{\text{3H-7H}} \rangle$  & 3.5(6) & 2.3(4) & 2.6(4) & 2.8(4) &
 3.0(4) & 2.9 \\
\end{tabular}
\end{ruledtabular}
\label{tab:3h-distr}
\end{table}


\subsection{Peak at 5.5\,MeV and possible 7.5\,MeV state}
\label{sec:exp-dis-6}


The peak in the $^{7}$H MM spectrum at $\sim 6$\,MeV was assumed in Ref.\
\cite{Bezbakh:2020} to correspond to the $5/2^+$--$3/2^+$ doublet or one of its
components.
For the discussion of this Section, we should assume (i) the possible width of
the state and (ii) the profile of the resonance peak, which is also induced by
this width.
The relevant theoretical estimates are provided in Ref.\ \cite{Grigorenko:2021}.
Contrary to the $^{7}$H g.s., which has a unique true $4n$ emission decay
channel, the components of the $5/2^+$--$3/2^+$ doublet, located above $E_T \sim
4$\,MeV, can undergo the sequential $^{7}$H $\rightarrow\, ^5$H(g.s.)+$2n$
$\rightarrow\,^3$H+$4n$ decay.
The alternative sequential decay channel via $^{6}$H is assumed to be closed,
because no $^{6}$H states available for the $^{7}$H$\rightarrow^6$H+$n$ decay
were found in this work below the 6\,MeV energy relative the $^3$H+$3n$
threshold.

\begin{figure}
\begin{center}
\includegraphics[width=0.47\textwidth]{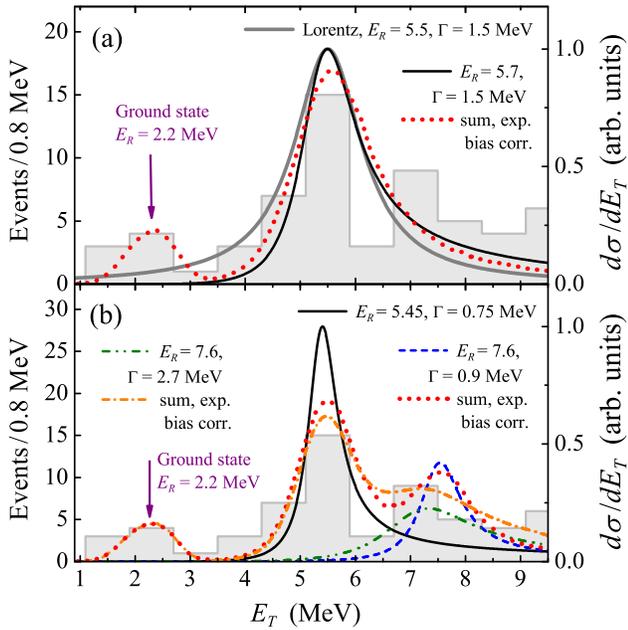}
\end{center}
\caption{
The energy profile of the ground and the first excited state(s) of $^{7}$H.
(a) Broad 5.5 MeV state case.
The gray solid curve shows pure Lorentzian profile with $\Gamma=1.5$\,MeV.
The black solid curve shows the realistic profile Eq.\ (\ref{eq:res-prof}) with
$\Gamma=1.5$\,MeV.
The red dotted curve shows the realistic resonance profile with experimental
resolution taken into account by the MC simulations.
(b) Narrow 5.5 MeV states case.
Black solid, blue dashed, and olive dash-double-dotted curves show the
$5/2^+$--$3/2^+$ doublet with the resonance energies of 5.5 and 7.5\,MeV, but
with different widths of the $3/2^+$ resonance state.
The red dotted and orange dash-dotted curves show the g.s.\ plus doublet spectra
with experimental resolution taken into account by the MC simulations.
The g.s.\ width is assumed to be very small ($\Gamma \lesssim 1$ keV) compared
to energy resolution.
The experimental $^{7}$H MM spectrum with gates $E_{3\text{H}}<4/7E_T$ and
$\theta_{\text{cm}}< 18^{\circ}$ is shown in both panels by gray histograms.
}
\label{fig:profile-for-simulations}
\end{figure}

We start with the overall ``pessimistic'' estimates for the resonance profile.
The upper-limit width value of the sequential decay of the ``$2^+$'' state at
5.5\,MeV via the $^{5}$H g.s.\ is determined in Ref.\ \cite{Grigorenko:2021} as
$\Gamma=0.75$\,MeV.
We assume a conservative value of $\Gamma=1.5$\,MeV.
It can be seen in Fig.\ \ref{fig:profile-for-simulations} (a) that Lorentzian
profile with such a width extends to the g.s.\ position and may ``shade'' it.
However, a realistic resonance profile should be corrected by the function of
the width dependence on energy:
\begin{equation}
\frac{d \sigma}{d E_T} \sim \frac{\Gamma(E_T)}{(E_r-E_T)^2+\Gamma(E_T)^2/4}\,.
\label{eq:res-prof}
\end{equation}
Functions of this type were proved to be extremely precise in the description of
the resonance profiles of three-body decays, see, e.g., Eq.\ (24) in Ref.\
\cite{Grigorenko:2009c} and  Eq.\ (7) in Ref.\ \cite{Grigorenko:2015b}.
So, we may expect that this profile is reasonably precise for the sequential
$(2n)-(2n)$ decay.
The low-energy width behavior of the ``$2^+$'' state should be described as
$$\Gamma(E_T)\sim E_T^4.$$
The difference with Eq.\ (\ref{eq:pv-0p}) should be emphasized, where the first
two neutrons can be emitted in the ``$0^+$'' configuration.
Let's assume the ``softer'' behavior $\Gamma(E_T) \sim E_T^3$ for simulations.
It can be found in Fig.\ \ref{fig:profile-for-simulations} (a) that even this
overall ``pessimistic'' estimate of the resonance profile results in a deep
minimum between the $^{7}$H ground and the first excited states.
For the narrower $^{7}$H excited state, see Fig.\
\ref{fig:profile-for-simulations} (b), the deep is much more pronounced.
So, a clear experimental separation of the $^{7}$H ground and first excited
states in our experiment is expected.

Additional clue on another resonant state can be found on the right tail of the
5.5\,MeV peak at about 7.5\,MeV in Fig.\ \ref{fig:rebin-test}, upper row.
It could be just a statistical fluctuation of the data.
However, one should keep in mind that, if the 5.5\,MeV peak is indeed the
$5/2^+$ state of $^{7}$H, than one may expect the $3/2^+$ member of this doublet
to be located $\sim 1-2$\,MeV above it.
According to statistical argument, the $3/2^+$ state population should be
1.5\,times smaller than the $5/2^+$ population.
For the $3/2^+$ state at $E_T\sim 7.5$\,MeV with a width smaller than
$1-1.5$\,MeV we can find that the doublet components can be resolved by the
setup of the experiment 2, see Fig.\ \ref{fig:profile-for-simulations} (b), red
dotted curve.
Otherwise, for the broad $3/2^+$ state, we can expect some quite broad
asymmetric ``triangular'' profile with ``shoulder'' for the $5/2^+$--$3/2^+$
doublet, see Fig.\ \ref{fig:profile-for-simulations} (b), the orange dash-dotted
curve.

So, we can not discriminate confidently among three possible interpretations of
the data around 7.5\,MeV: (i) contributions from asymmetric broad right
``shoulder'' of the 5.5\,MeV state [Fig.\ \ref{fig:profile-for-simulations}
(a)], (ii) two broad overlapping states, and (iii) two narrow distinguishable
states [Fig.\ \ref{fig:profile-for-simulations} (b)].
However, the idea about contribution from the $3/2^+$ doublet state is
attractive, since it allows also to explain some difference with the experiment
1.
The $E_T=6.5(5)$\,MeV peak position found in \cite{Bezbakh:2020} is consistent
with the observation anticipated for the unresolved $5/2^+$--$3/2^+$ doublet
$E_T=\{5.5,7.5\}$\,MeV.
It can be found in Fig.\ \ref{fig:rebin-test} (e) that a spectrum, consistent
with the results of experiment 1, can be obtained with a large bin size.


\subsection{Group of events at 11 MeV}
\label{sec:exp-dis-11-14}


The 11.0(5)\,MeV peak is well seen in all the data representations in Fig.\
\ref{fig:exp-2-all} (b,c) and \ref{fig:rebin-test} (a-e).
The search for this state at $\theta_{\text{cm}} > 18^{\circ}$ loses sense,
since this energy-angular range is expected to be contaminated with the
background events, see Fig.\ \ref{fig:exp-empty} (b).

Question could be asked: what could be the nature of such quite a narrow states
observed at such a high excitation energy?
The disintegration of the $^{3}$H cluster into $p$+$n$+$n$ is possible above
$E_T=8.48$\,MeV.
Phenomenology of nuclear states suggests that the states with definite
clusterization are likely to be found near the corresponding cluster
disintegration thresholds (both somewhat above and somewhat below).
According to this systematics the 11\,MeV state can be expected to have the
structure with ``dissolved'' $^{3}$H core: $p$+$6n$.
It does not mean that such a state should be necessarily observed in the
$p$+$6n$ decay channel.
According to a penetrability argument, the $^{3}$H+$4n$ channel should still be
a preferable decay path for such a state.
Nevertheless, we performed a dedicated search for decay of this state into
$p$+$6n$.
Unfortunately, no significant concentration of such events was identified.


\subsection{Neutron coincidence events}
\label{sec:exp-dis-neutrons}


The neutron wall used in experiment 2 provided $4.5 \%$ energy resolution and
the single neutron registration efficiency of $\sim 15 \%$.
The efficiency of the neutron registration in coincidence with $^{3}$H and
$^{3}$He was around $2 \%$ taking into account that four neutrons are produced
in each $^{7}$H decay event.
Such an efficiency is too low to expect statistically significant result.
However, these events could be interesting as an additional consistency check of
the data, see Fig.\ \ref{fig:exp-2-all} (a) and (c).
There are 8 triple-coincidence $^{3}$H-$^{3}$He-$n$ events.
There are two events in the $E_T\sim 5.5$\,MeV region, one event corresponding
to $\sim 7.5$\,MeV structure, and three events consistent with the 11\,MeV
state.
This is an encouraging result, since $\sim 75 \%$ of neutron events coincide
with expected regions for resonance states, while only $\sim 20\%$ of the data
is concentrated on these regions.


\subsection{$^{7}$H c.m.\ angular distributions}
\label{sec:exp-angdis}


The c.m.\ angular distributions of the direct reactions serve as one of standard
tools of spin-parity identification in reaction theory.
Due to the small statistics of our experiment, angular distributions can not
provide a basis for reliable statements, but some conclusive remarks still can
be done.

Fig.\ \ref{fig:exp-empty} (c) shows the center-of-mass angle of events from the
$^2$H($^8$He,$^3$He)$^{7}$H reaction versus the corresponding $^{7}$H MM energy
taken from the experiment 2.
For $E_T> 10$\,MeV the available angular range rapidly shrinks: the kinematical
cut defined by the maximum energy $E=26$\,MeV of reliable identification of
$^{3}$He recoils can be clearly seen.
The angular distributions for different energy ranges of $^{7}$H are presented
in Fig.\ \ref{fig:angdis}. The efficiency-corrected angular distributions
converted to cross section values are shown in the upper panels of Fig.\
\ref{fig:angdis}.

The angular distribution for the possible $^{7}$H g.s.\ energy range, shown in
Fig.\ \ref{fig:angdis} (a), has already been discussed in Section
\ref{sec:exp-angdis-gs}.
Here we would like to point the deduced cross sections are $\sim 24$ $\mu$b/sr
for $\theta_{\text{cm}} \sim 5^{\circ}-9^{\circ}$ and $\sim 7$ $\mu$b/sr for
$\theta_{\text{cm}} \sim 15^{\circ}-19^{\circ}$.
For the first excited state, Fig.\ \ref{fig:angdis} (b), the deduced cross
sections are $\sim 30$ $\mu$b/sr for $\theta_{\text{cm}} \sim
5^{\circ}-18^{\circ}$ and $\sim 11$ $\mu$b/sr for $\theta_{\text{cm}} \sim
18^{\circ}-30^{\circ}$.
The energy range $8.5<E_T<10.5$\,MeV was excluded from consideration, because of
the remarkable background found in the empty target experiment, see Fig.\
\ref{fig:exp-empty} (b).

In general, we would like to comment the following.
Within the available angular range and available statistics, the angular
distributions of all the 4 ranges can be seen as qualitatively different.
Thus, these distributions support the idea that the considered ranges contain
physically different entities.


\section{Conclusions}


In this work we provide extended discussion of the $^{7}$H data obtained for the
$^2$H($^8$He,$^3$He)$^{7}$H reaction in the experiment Ref.\ \cite{Bezbakh:2020}
and the new data for the same reaction but with an improved setup.
The statistics collected in the last experiment ($378$ events) is considerably
larger than in \cite{Bezbakh:2020} ($119$ events) and in Ref.\
\cite{Nikolskii:2010} ($\sim 100$ events).
In the experiment 2 \emph{four} peaks are observed in the MM spectrum of $^{7}$H
at $2.2(5)$, $5.5(3)$, $7.5(3)$, and $11.0(3)$\,MeV.
This result is consistent with \emph{three} bumps in the spectrum observed at
$\sim 2-3$, $\sim 6$, and $\sim 11$\,MeV in the experiments \cite{Bezbakh:2020}
and  \cite{Nikolskii:2010}.
For each of these three datasets, because of the limited statistics, it is not
impossible that these peaks are induced by statistical fluctuations on a top of
some smooth continuous spectrum starting from $E_T \sim 5$\,MeV.
However, it is virtually impossible for statistical fluctuations to cause peaks
at the same energies in the three totally independent experiments.

The $^{7}$H g.s.\ is extremely poorly populated in the
$^2$H($^8$He,$^3$He)$^{7}$H reaction.
Possible reasons for this suppression are presumably connected with very
radially extended and, thus, very ``fragile'' nature of this state; the issue is
separately discussed in the forthcoming theoretical article
\cite{Grigorenko:2021}.
So, our special concern in this work were the background conditions in the
low-energy part of the spectrum and the energy-resolution issues, which may make
possible ``contamination'' of the g.s.\ range by events from the higher-lying
$^{7}$H excitations.
Both these aspects were found to be favorable for the $^{7}$H g.s.\
identification even by few events.
The 5 $^{7}$H g.s.\ candidate events were collected in the experiment 1
\cite{Bezbakh:2020} and 9 events in the experiment 2.
All the observed events are consistent with the $^{7}$H center-of-mass angular
distribution expected for the $1/2^+$ g.s.\ with diffraction minimum located
between $\sim 10^{\circ}$ and $\sim 15^{\circ}$.
They are also consistent with predicted energy distributions of the $^{3}$H
fragment in the $^{7}$H center-of-mass system.

Summarizing, the conclusion about the observation of the $^{7}$H states at
$2.2(5)$ and $5.5(3)$\,MeV is very reliable.
The observation of the $7.5(3)$\,MeV state is not statistically confident
enough.
Energy resolution of the experiment 2 was high enough to resolve the possible
$5.5$ -- $7.5$\,MeV doublet (while in the experiments \cite{Bezbakh:2020} and
\cite{Nikolskii:2010} they were observed as a single structure).
However, we can not exclude that the observed separation of the $5.5$ --
$7.5$\,MeV peaks is actually a statistical fluctuation on the broad right tail
of the $5.5$\,MeV state.
Anyway, we conclude that the firmly ascertained 5.5(3)\,MeV state is the $5/2^+$
member of the $^{7}$H excitation doublet.
The 11\,MeV peak is well exhibited at low center-of-mass angles
$\theta_{\text{cm}} \lesssim 20^{\circ}$, where available statistics is limited.
It is also well seen at higher center-of-mass angles $\theta_{\text{cm}} \sim
20^{\circ}-35^{\circ}$.
However, in this energy-angular range a strong background contribution is
expected, so caution is needed.


\acknowledgments


We acknowledge the principal support of this work by the Russian Science
Foundation grant No.\ 17-12-01367.
The authors are grateful to Profs.\ Yu.Ts.\ Oganessian and S.N.\ Dmitriev for
the long-term support and development of this activity.
We acknowledge important contribution of Prof.\ M.S.\ Golovkov to the
development of the experimental methods and useful discussions.
Also, authors express their gratitude to the acceleration team for the stable
work of U-400M cyclotron during all runs.


\bibliographystyle{apsrev4-2}
\bibliography{all}


\end{document}